\documentclass[preprint2]{aastex}

\input epsf

\received{February 7, 2001}
\accepted{March 30, 2001}

\slugcomment{To Appear in the Astrophysical Journal, Volume 556, July 20, 2001}

\shorttitle{XTE J1118+480}
\shortauthors{Wagner et al.}

\newcommand{\kms}{\ifmmode {\rm km\ s}^{-1} \else km s$^{-1}$\fi}
\newcommand{\Msun}{\ifmmode {\rm M}_{\odot} \else M$_{\odot}$\fi}
\newcommand{\Av}{\ifmmode A_{\rm V} \else $A_{\rm V}$\fi}
\newcommand{\Mv}{\ifmmode M_{\rm V} \else $M_{\rm V}$\fi}
\newcommand{\mv}{\ifmmode m_{\rm V} \else $m_{\rm V}$\fi}
\newcommand{\iauc}{IAU Circ.~}
\newcommand{\iraf}{\sc{iraf}\rm}

\begin{document}

\title{The Halo Black-Hole X-ray Transient 
XTE~J1118+480\altaffilmark{1}}


\author{R. Mark Wagner\altaffilmark{2}, C. B. Foltz\altaffilmark{3}, T.
Shahbaz\altaffilmark{4}, J. Casares\altaffilmark{4}, P. A.
Charles\altaffilmark{5},\\ S. G. Starrfield\altaffilmark{6}, \& P.
Hewett\altaffilmark{7}}


\altaffiltext{1}{Based in part on observations obtained at the MMT
Observatory, a joint facility of the University of Arizona and the Smithsonian
Institution.}  \altaffiltext{2}{Large Binocular Telescope Observatory,
University of Arizona, Tucson, Arizona 85721; rmw@as.arizona.edu}   
\altaffiltext{3}{MMT Observatory, University of Arizona, Tucson, Arizona 
85721; cfoltz@as.arizona.edu} 
\altaffiltext{4}{Instituto de Astrof\'\i{}sica de Canarias, 38200 La Laguna,
Tenerife, Spain; tsh@ll.iac.es, jcv@ll.iac.es}  
\altaffiltext{5}{Department of Physics \& Astronomy, University of
Southampton, Southampton, SO17 1BJ, UK; pac@astro.soton.ac.uk}
\altaffiltext{6}{Department of Physics and Astronomy,  Arizona State University,
Tempe, Arizona 85287; sumner.starrfield@asu.edu}  
\altaffiltext{7}{Institute of Astronomy, Cambridge University, Madingley Road,
Cambridge, CB3 0HA, UK; phewett@ast.cam.ac.uk} 


\begin{abstract}
Optical spectra were obtained 
of the optical counterpart of the high latitude ($b\simeq62\degr$)
soft X-ray transient XTE J1118+480 near its quiescent state ($R \simeq
18.3$) with the new 6.5~m MMT and the 4.2~m WHT.  The spectrum exhibits broad,
double-peaked, emission lines of hydrogen (FWHM $\simeq$
2400~\kms) arising from an accretion disk superposed with absorption lines of a
late-type secondary star.  Cross-correlation of the 27 individual
spectra with late-type stellar template spectra reveals a sinusoidal
variation in radial velocity with amplitude $K = 701 \pm 10\ \kms$
and orbital period $P = 0.169930 \pm 0.000004$ d. The mass function,
$6.1 \pm 0.3\ \Msun$, is a firm lower limit on the mass of the compact
object and strongly implies that it is a black hole.  We
estimate the spectral type of the secondary to be K7V--M0V and that it
contributes $28 \pm 2\%$ of the light in the 5800-6400 \AA\ region on 2000
November 20 increasing to $36 \pm 2\%$ by 2001 January 4 as the disk fades.
Photometric observations ($R$-band) with the IAC 0.8~m telescope
reveal ellipsoidal light variations of full amplitude $0.2$ mag.
Modeling of the light curve gives a large mass ratio ($\frac{M_1}{M_2}
\sim 20$) and a high orbital inclination ($i = 81\degr \pm 2\degr$). 
Our combined fits yield a mass of the black hole in the range $M_1 =
6.0-7.7~\Msun$ (90\% confidence) for plausible secondary star masses
of $M_2 = 0.09-0.5~\Msun$.  The photometric period measured during the
outburst is 0.5\% longer than our orbital period and probably reflects
superhump modulations as observed in some other soft X-ray transients.  The 
estimated distance is $d = 1.9 \pm 0.4$ kpc corresponding
to a height of $1.7 \pm 0.4$ kpc above the Galactic plane.  The spectroscopic,
photometric, and dynamical results indicate that XTE J1118+480 is the first
firmly identified black hole X-ray system in the Galactic halo.

\end{abstract} 




\section{Introduction}

X-ray novae or soft X-ray transients comprise a subset of low-mass
X-ray binaries (LMXBs) which consist of a late-type secondary star and
a neutron star or a black hole exhibiting bright optical and X-ray
outbursts which are recurrent on time scales of decades
\citep{tanaka95, tanaka96}.  During their outbursts, they resemble
persistent LMXBs in which the light of the secondary star is
overwhelmed by a luminous accretion disk surrounding the compact
object.  After a year or less in some objects, the system returns to
quiescence.  The secondary star now contributes a much larger
fraction of the total light and its atmospheric absorption lines become
visible in optical spectra.  Thus quiescent X-ray novae provide
the ideal opportunity to study the nature and dynamical properties of
the binary system (see e.g. Charles 1998).  These studies have
demonstrated that the mass of the compact object in ten X-ray novae
(Charles 1998; McClintock 1998; Nova Vel 1993: Filippenko et al. 1999;
V4641 Sgr: Orosz et al. 2001) exceeds the theoretical maximum mass of
a neutron star \citep{rhoades74} and thus must evidently be a black hole.

A previously unknown X-ray transient, XTE J1118+480, was discovered by the
RXTE ASM instrument on 2000 March 29 \citep{remillard00} with an average 2-12
keV X-ray intensity of only 39 mCrab, but with a spectral signature
similar to other accreting Galactic black holes.  An optical
counterpart was identified on March 30 \citep{uemura00} with $V =
12.9$ and confirmed spectroscopically \citep{garcia00}.  Photometry
during the outburst \citep{patterson00,uemura00} revealed a weak
modulation with a period of$\simeq 0.1708$ d and an amplitude of
$0.055$ mag reflecting the probable orbital period of the system.  The shape
of the light curve and its temporal evolution resembled those of superhumps
\citep{uemura00} observed during superoutbursts of short-period
cataclysmic variables and outbursts of some other soft X-ray transients
\citep{odonoghue96}, and which correspond to a period only fractionally 
($\leq$1--2\%) longer than the orbital period.

By 2000 August 31, photometry reported to VSNET indicated that
J1118+480 had faded to near its quiescent level of $V \simeq 18.8$.
Its high Galactic latitude ($b = +62\degr$), implying a low value of
interstellar absorption (\Av = 0.04 mag; Hynes et al. 2000), combined with its
relatively bright quiescent magnitude makes J1118+480 ideal for detailed study
in quiescence, especially at wavelengths not easily observable in other
low-latitude transients.  In this {\it Letter}, we present the results
of optical spectroscopy and photometry obtained in quiescence
beginning in late 2000 November and continuing through 2001 January.
A preliminary announcement of our spectroscopic results
\citep{wagner00} as well as those obtained by a second group
\citep{mcclintock00} was reported in the IAU Circulars.

\section{Observations}

We obtained 20 spectra of J1118+480 with the newly upgraded 6.5~m MMT
equipped with the blue channel CCD spectrograph on the nights of 2000
November 20 (5 spectra), 30 (9 spectra), and 2001 January 4 (6
spectra). The spectra cover the range $\lambda\lambda$4200-7500 at a
spectral resolution of 3.5 \AA\ and a dispersion of 1.1 \AA\
pixel$^{-1}$. The seeing was typically 1\arcsec~and a 1\arcsec~wide
slit was employed.  Each 1440 s exposure was bracketed by a HeNeAr
lamp spectrum which led to a wavelength calibration accurate to 5--7
\kms\ rms.  Spectra of the late-type
dwarf stars BD+63\degr 137 (K7), Gliese 239 (M0), Gliese 96 (M0.5),
Gliese 154 (M0), and Gliese 388 (M3.5) were obtained using the same instrumental
configuration so as to accurately gauge the spectral type of the
secondary star and to provide templates for the radial velocity
analysis.  We note that the first two stars, BD+63\degr 137 and Gliese
239, are classified as K5V and K7 respectively in Simbad.  However,
based on the depth of their TiO bands and by comparison with very
securely classified templates (such as 61 Cyg A \& B), we support
classifications of K7V and M0V respectively.
 
We also obtained 7 spectra of J1118+480 with the ISIS red channel of
the 4.2~m William Herschel Telescope on La Palma on the night of 2001
January 11. We employed the R316R grating which yields 1.47 \AA\
pixel$^{-1}$ in the range $\lambda\lambda$5820-7320. The slit width
was $1\farcs5$ which resulted in a spectral resolution of 4.5
\AA. Exposure times were 1200 s and the wavelength calibration
provided by internal lamps was checked with respect to night-sky
emission lines to be within 10 \kms.  Spectra of the dwarf stars HR
5265 (K3), HR 8085 (61 Cyg B; K5), HR 8086 (61 Cyg A; K7), and Gliese
361 (M1.5) were both employed as radial velocity templates and for spectral
classification of the secondary.  Accurate absolute photometric
calibration was not attempted for either the MMT or WHT spectra due to
slit losses.

Time-resolved differential CCD photometry was obtained of J1118+480 in
the $R$-band ($\lambda_c$ = 6500 \AA; FWHM = 1500 \AA) on the nights of
2000 December 14, 28, and 2001 January 9 (UT) with the 0.80~m IAC80
telescope at the Observatorio del Teide (Tenerife).  We used the
Thompson $1024 \times 1024$ pixel CCD camera with $2 \times 2$ pixel binning
giving an image scale of $0\farcs43$ pixel$^{-1}$.  A total of 68
10-minute exposures were obtained in $\sim2\arcsec$ seeing and good
transparency, together with bias and flatfield frames to facilitate
the standard data reduction procedures.  We applied profile fitting
photometry to J1118+480 and several nearby comparison stars using
\iraf. We also selected comparison stars which were checked
for variability during the night and calibrated as field reference
stars using Landolt standard stars to place our photometry on an
absolute scale.  We estimate that our relative photometric accuracy is
0.05 mag.  Finally, we measured the position of J1118+480 on a registered
average of the frames obtained on December 28.  Based on a grid of 22 USNO A2.0
stars, we find that the position of J1118+480 is R.A. = $11^h18^m10\fs84$,
Decl. = +$48\degr02\arcmin12\farcs9$ (equinox 2000.0, accuracy $\pm 0\farcs3$).

\section{Results}
\subsection{Spectroscopy}

The individual spectra of J1118+480 exhibit broad Balmer emission
lines arising from an accretion disk superposed with absorption lines
or bands of Mg{\it b} $\lambda5175$, Na D, and of TiO.
Radial velocities were extracted by cross-correlating the individual
normalized spectra over the range 5800-6400 \AA\ of J1118+480 with our grid of
templates using routines in both \iraf\ and locally developed software
packages.  Phasing of the radial velocities on the photometric period
(0.17 d) yielded a nearly sinusoidal variation in radial velocity indicating
that the photometric period was close to the orbital period.  

To obtain an accurate measurement of the
orbital period and determine the best-fitting spectral type of the secondary,
successive circular orbits with periods ranging from 0.169000 to 0.171000 d
were fit to the radial velocity data defined by the grid of templates. We
found that our best cross-correlations and circular orbit solutions
corresponded to spectral types of K5V-M0V and an
orbital period of $0.169930 \pm 0.000004$ d ($\chi_\nu^2 = 4.5;
23$ dof), i.e. 0.5\% shorter than
the outburst photometric period of Uemura et al. (2000). We have scaled the
velocity errors by a factor of $\sqrt{4.5}$ in order to give $\chi_\nu^2 =
1.0$.  

The radial velocity curve of the secondary star is shown in Figure 1
(upper panel).  A sine fit to the data gives: $P = 0.169930 \pm
0.000004$\ d; $\gamma = -15\pm 10\ \kms$; $K = 701
\pm 10\ \kms; T_0 = HJD\ 2,451,868.8916 \pm 0.0004$ (2000 November 20.3916 UT),
where $T_0$ corresponds to absolute phase 0.0 or inferior conjunction of the
secondary star.  Orbital smearing will systematically underestimate the true
velocity amplitude.  For these data at the peak of the radial velocity curve, we
estimate this effect to be no more than 1.6\% or about 11 \kms.  The mass 
function is

\begin{displaymath} f(M) =
\frac{(M_1 \sin i)^3}{(M_1 + M_2)^2} = \frac{PK^3}{2\pi G} = 6.1 \pm 0.3\
M_\sun, \end{displaymath}

\noindent where G is the Gravitational Constant.  The mass function implies a
lower limit to $M_1$ of $M_1 = 6.1\ M_\sun$.  This greatly exceeds the maximum
mass of a neutron star \citep{rhoades74} and implies that the compact object is
probably a black hole.

In Figure 2, we show the spectrum of J1118+480 on 2000 November 30,
formed by averaging the individual Doppler-corrected spectra (total
integration time 3.6 hr) to the rest frame of the secondary star,
together with the spectra of a M0.5V (Gliese 96) and K7V (BD+63~137)
star for comparison, bracketing the likely spectral type of the
J1118+480 secondary.  The absorption features of the secondary star are
now sharp and readily apparent, but this process smears
out the emission lines arising from the accretion disk surrounding the
compact object.  However, comparison of the spectrum with our
templates indicates that the spectrum of J1118+480, apart from the
emission lines, is too blue.  Also, the absorption lines are much
weaker relative to an uncontaminated K7V star, thereby suggesting a
substantial contribution from the quiescent accretion disk.  This
contribution, particularly the velocity information in the broad and
complex H$\alpha$ emission line, will be discussed in a forthcoming
companion paper \cite{zurita01}.

To estimate the fraction of the observed flux arising from the secondary
star and disk separately, we subtracted fractions of our grid of templates
from the Doppler-corrected spectra of J1118+480 for each night and
performed a $\chi^2$ minimization of the residuals in the range 5800-6400 \AA.  
This region overlaps both the MMT and WHT spectra and is rich in metallic
absorption lines. The templates were broadened by convolution with
Gaussian passbands to account for small differences in instrumental
resolution between the MMT and WHT and the effects of orbital smearing due
to the length of our exposures. We find that the best-fitting templates
are K7V-M0V and that the secondary contributes $\simeq32\%$ of the total
light in this spectral region. Our results are consistent with those
obtained by McClintock et al. (2001) close to our earlier observations.
Specifically, we find that for 2000 November 20, 30, and 2001 January 4
that the secondary contributes $28 \pm 2\%$, $28 \pm 2\%$, $36 \pm
2\%$ respectively, so that it appears that the disk luminosity is still
declining and the system may not yet be fully quiescent.

\subsection{Photometry}

To interpret the optical light curves, we used a model that includes a
Roche lobe-filling secondary star, a concave accretion disk, and mutual
eclipses of the disk and the secondary star. This X-ray binary model
and the fitting procedure are fully described in Shahbaz et al.
(2001). The model parameters are the binary inclination $i$, the mass
ratio $q$ (=$M_1/M_2$), the mean temperature $\bar{T}$ and gravity
$\log \bar{g}$ of the secondary, the gravity darkening exponent
$\beta$, the accretion disk radius $R_{disk}$ [defined as the fraction
of the distance to the inner Lagrangian point ($R_{L_1}$)], the flare
angle of the accretion disk ($\theta$), the temperature at the outer disk
edge $T_{disk}$, and the exponent on the power-law radial temperature
distribution $\eta$.  We fixed $\Av$ at 0.04 mag \citep{hynes00}.

From our optical spectroscopy, the secondary star has an observed
spectral type of K7V-M0V, so we fixed $\bar{T}$ at 4250~K and $\log
\bar{g}$ at 5.0, appropriate for K7V. Since the late-type star is convective,
we take $\beta=0.08$ \citep{lucy67}.  The accretion disk is assumed to
be optically thick and geometrically thin, and so we set the flare
angle of the disk to be 0.04 radians and $\eta$ to be -0.75
(appropriate for a steady-state disk; Pringle 1981). We can use the
value for the mass function and the observed spectral type to place an
upper limit to the binary mass ratio. The mass function gives a lower
limit to $M_1$ ($\geq6.1~M_{\odot}$) and the mass for the main
sequence spectral type of the secondary gives an upper limit for $M_2$
($\la0.52~M_{\odot}$). Thus we obtain a lower limit for the mass ratio
of $q\ga12$ and so we assume $q = 20$ in our model (the results are
largely insensitive to this value).  Note that such high mass ratios
are also supported by our small ``superhump period'' excess in the
model of Mineshige et al. (1992).

The phase-averaged $R$-band light curve of J1118+40 is shown in the
lower panel of Figure 1 together with our best-fitting model for the
parameters $i$, $R_{disk}$, $T_{disk}$, a phase
shift,the distance $d$, and normalization. The data were folded
using the spectroscopic period and epoch given above (i.e. phase 0 is
defined as superior conjunction of the compact object; note that this
is {\it not} how McClintock et al. (2001) have defined their phase
convention).  We find a best fit at $i=81 \pm 2$ degrees,
$R_{disk}=0.8R_{L1}$, T$_{disk} = 3906$~K, and $d = 1.8
\pm 0.3$ kpc (2$\sigma$).  A one--dimensional grid search was
performed to determine the uncertainties of the fitted parameters.
The model predicts a veiling of 76\% which is consistent with the
values derived from our optical spectroscopy.  It should be noted that
with such a high inclination one might normally expect to see eclipses
of the secondary and accretion disk.  However, given the extreme mass
ratio, one only expects shallow eclipses of the secondary and disk (Figure 1).

In addition, we find a small phase shift ($0.02 \pm 0.04$;
$\sim$4.6 min) between the photometric light curve and the
spectroscopically defined phase.  Such an effect (at the level of 12
$\pm$ 2 min) is seen by McClintock et al. (2001) who, as we do,
discount the possibility of an instrumental origin.  It should be
noted that McClintock et al. do not formally fit their data with an
ellipsoidal model, but show one that ``matches'' their data, albeit
with a very obvious phase offset of 12 min.  Their model combines an
ellipsoidal modulation with a constant contribution (at the 66\%
level) from an accretion disk. We believe that this phase shift is
actually due to a temporally varying, {\it asymmetrically} emitting
disk component.  In a companion paper (Shahbaz et al. 2001), we model
the quiescent light curves of the neutron star SXT XTE~J2123-058 (it
has a 6 hr orbital period and high orbital inclination) which
are remarkably similar in shape to that of McClintock et al. (2001)
by employing such a disk component.

\section{Discussion}

The mass of the secondary star can be constrained by making
assumptions regarding its evolutionary state. An upper limit to $M_2$
can be obtained by using the mass of a main sequence star of the same
spectral type ($M_2(MS)$), while the secondary would not have left the
main sequence if its mass were less than the Schonberg-Chandrasekhar
limiting value of 0.17$M_2(MS)$. Thus from the observed spectral type
of K7V, $M_2$ must lie in the range 0.09--0.50~$M_{\odot}$.  Combining
this constraint with our measured values for $f(M)$ and $i$, we show
in Figure 3 the inferred range for $M_1$ which is 6.0--7.7~$M_{\odot}$
(2$\sigma$).  The uncertainties were determined by Monte Carlo
simulation, in which we draw Gaussian--distributed random values for
the observed quantities, with a mean and variance the same as the
observed values.

Over the past 10 years dynamical studies have 
established that several X-ray novae contain compact objects with 
masses significantly larger than the maximum mass of a normal neutron star
($\simeq3.2$\ \Msun; Rhoades \& Ruffini 1974) such as V404 Cyg ($\ga 6.1\
M_\sun$; Casares and Charles 1994); QZ Vul 1988 ($\ga 5.0\ M_\sun$; Casares,
Charles, \& Marsh 1995; Filippenko, Matheson, \& Barth 1995; Harlaftis, Horne,
\& Filippenko 1996); and V2107 Oph 1977 ($\ga 4.7\ \Msun$; Remillard et al.
1996; Filippenko et al. 1997; Harlaftis et al. 1997) implying that
they are black holes. The large mass function and well-determined mass
of the compact object in XTE J1118+480 derived here indicates that it
almost certainly contains a black hole rather than a neutron star.
Recently Bailyn et al. (1998) have noted that the masses of black
holes in LMXBs are strongly peaked at $\sim$7\Msun, and our mass
determination is consistent with this result.

We can estimate the distance to J1118+480 by constraining the
secondary's size from our photometric modeling by using the Eaton \& Poe (1984)
redetermination of the Barnes-Evans relation, which relates $V$-band flux at the
star and color for stars with measured angular diameters.  From our light curve
the mean $R$ is 18.17, of which 76\% is from the disk according to our model
fits.  This implies that the secondary has $R$ = 19.72, and if its spectral
type is K7V ($V-R$ = 1.15; from our spectroscopy), then $V_0$ = 20.83 (assuming
\Av = 0.04).  From $V_0$ and the $V-R$ color, we then
estimate that the angular diameter of the secondary is $1.94 \times
10^{-3}$ mas \cite{eaton84}.  From the Eggleton (1983)
approximation of the volume radius of the Roche lobe of the secondary star, we
find that the size of the Roche-lobe is $R_2 = 0.17a$ for $q = 20$ and where $a$
is the separation of the center of mass of the two components. From our total
mass estimate of $7\ \Msun$ and Kepler's third law, we find that $a = 2.37
R_\sun$, and thus $d = 1.9 \pm 0.4$ kpc.  The error in $d$ is dominated by the
uncertainty in the secondary spectral type.  We find that for $b = +62\degr$,
$z = 1.7 \pm 0.4$ kpc, where z is the vertical distance above the Galactic
plane.

Thus, XTE~J1118+480 is the first firmly identified black hole X-ray
binary system in the Galactic halo.  This fact is quite remarkable
compared with the distribution of the black-hole LMXBs \citep{white95}
where the mean $z$ is around 400~pc (and that is dominated by a single
object, H1705-25, at $\sim$950~pc).  This is less
than half that of the neutron star systems, which is explained by the
additional ``kick'' velocity received in the formation of the neutron
star.  To place J1118+480 at its current location in this scenario would have
required an enormous velocity out of the plane, making its almost zero systemic
velocity very unlikely.  Furthermore, we find no evidence of a
large proper motion of J1118+480 ($\la10$ mas yr$^{-1}; 2\sigma$) since its
position measured during the outburst \citep{uemura00,masi00} and our
position near quiescence (see \S2) as compared with
its position measured on an archival POSS-I plate obtained $\sim47$ yrs ago
(from the USNO A2.0 and APM Palomar Schmidt sky catalog; Lewis \& Irwin 1996)
all agree to within the astrometric errors. This object therefore presents
interesting challenges and constraints on the formation and evolution of SXTs.

It is also interesting to consider whether there might be an
additional population of such high $z$ BH LMXBs in the Galaxy.
J1118+480 is one of the closest, yet is also one of the
(intrinsically) X-ray weakest of the LMXB X-ray transients and has a
low $L_X/L_{opt}$ ratio.  Hence, further examples would almost
certainly have gone undetected due to their X-ray faintness and high
latitude (the most sensitive X-ray monitoring is undertaken in the
plane and around the Galactic center).  However, Hynes et al. (2000)
discussed the possibility that the intrinsic weakness of the X-rays
relative to the optical might be due to the high inclination, making
this system akin to the Galactic ADC (accretion disk corona) sources,
and hence other high latitude systems would be expected to be much
brighter X-ray sources.  While this appears to have been borne out by
our inference of the (extremely) high inclination of 81\degr, there
are several difficulties with this argument.  There was an absence of any
X-ray modulation in outburst, although the secondary is so small that
it might always be completely shadowed by the disk and hence unable to
produce any significant modulation. (Note that this interpretation is
testable via higher time resolution photometry as the eclipse should
be visible; see Figure 1 at phase 0.0).  More seriously, the models
considered by Esin et al. (2001) and fit to the exceptionally wide
wavelength spectra of J1118+480 obtained during outburst have ruled
out the ADC model since the source was in the low/hard state.  This is
further supported by the direct X-ray observation of $\sim$0.1Hz QPOs
\cite{wood00}.


\acknowledgments

CBF and RMW would like to generously thank the staffs of the Steward
Observatory, Smithsonian Astrophysical Observatory, \& the MMT Observatory for
their extraordinary work and effort that has made the 6.5~m MMT a reality.  
The WHT/IAC80 telescopes are operated on the island of La Palma/Tenerife
by the ING/IAC in the Spanish Observatorio del Roque de Los
Muchachos/Teide.  We
are particularly grateful to Pablo Rodriguez-Gil and the IAC team that undertake
the Service Programme at Izana for obtaining some of the lightcurves that were
used in this analysis.  We also thank R. Corradi for taking the WHT spectra. 
PAC thanks Rob Hynes and Philipp Podsiadlowski for
useful discussions.  The authors wish to thank Felix Mirabel and Alex
Filippenko for constructive comments and discussions which greatly improved
the paper.  CBF acknowledges the support of NSF grant AST 98-03072.  SGS
acknowledges partial support by NSF and NASA grants to Arizona State
University.  RMW acknowledges partial support by NASA to The Ohio State 
University.

\begin{figure}
\epsfxsize = \hsize
\epsfbox{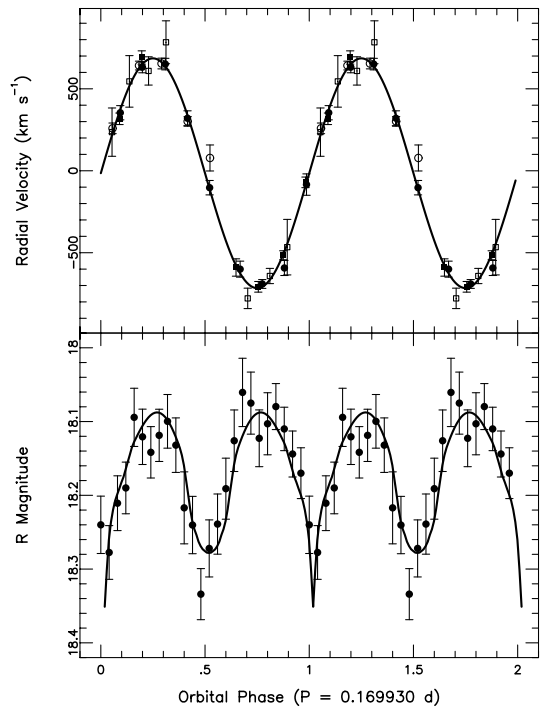}
\caption{Upper: Folded radial velocities of XTE J1118+480 and
the best-fitting  sinusoid.  Open circles indicate the data obtained on 2000
November 20 (MMT), filled circles on 2000 November  30 (MMT), filled squares on
2001 January 4 (MMT), and open squares on January 12 (WHT) respectively. Lower:
The phase folded $R$-band light curve of XTE J1118+480. The solid  line shows
the best--fit light curve solution. The sharp dip at phase 1.0 might be
due to a grazing eclipse of the accretion disk by the secondary star.  Two
orbital cycles are shown for clarity.}  
\end{figure} 

\clearpage

\begin{figure}
\epsfxsize=\hsize
\epsfbox{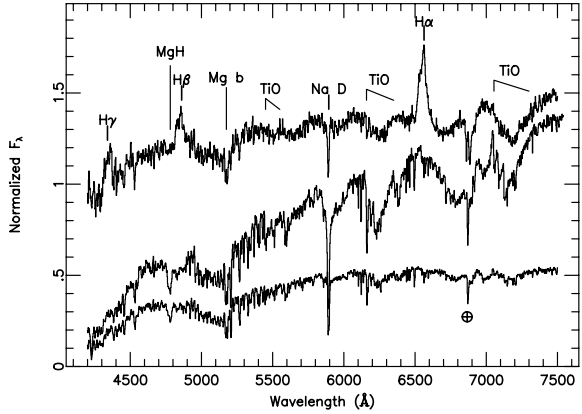}
\caption{Upper: Doppler-corrected rest-frame spectrum of XTE
J1118+480 obtained on 2000 November 30.  Middle: M0.5V spectrum (Gliese 96). 
Lower: K7V spectrum (BD+63\degr 137).  We estimate that the secondary is K7V-M0V
in J1118+480.  The  spectra have been offset for clarity.  The prominent
spectral features are indicated.}
\end{figure}   


\begin{figure}
\epsfxsize=\hsize
\epsfbox{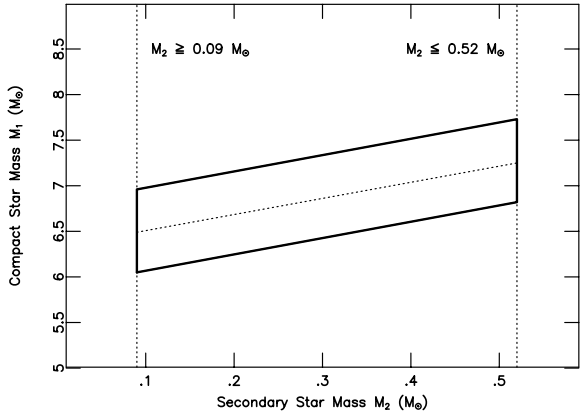}
\caption{The mass of the compact object as a function of the
secondary star mass. The two solid lines represent the 90\% confidence
limits for the probable mass range (the central dashed line being the
best fit value). The two dashed vertical lines are upper and lower limits to the
secondary star mass from other considerations (see text).}
\end{figure}


\end{document}